\documentclass[preprint2]{aastex}

\shorttitle{Outflow in Dust Disks}
\shortauthors{Takeuchi \& Lin}

\begin{document}


\title{Surface Outflow in Optically Thick Dust Disks by the Radiation
Pressure \footnote{To appear in Aug. 10 issue of Astrophys. J.}}

\author{Taku Takeuchi and D. N. C. Lin}
\affil{UCO/Lick Observatory, University of California, Santa Cruz,
CA95064}
\email{taku@ucolick.org, lin@ucolick.org}

\begin{abstract}
We study the outflow of dust particles on the surface layers of optically
thick disks. At the surface of disks around young stars, small dust
particles (size $\la 10 \ \micron$) experience stellar radiation
pressure support and orbit more slowly than the surrounding gas.
The resulting tail-wind imparts energy and angular momentum to the dust
particles, moving them outward.
This outflow occurs in the thin surface layer of the
disk that is exposed to starlight, and the outward mass flux is carried
primarily by particles of size $\sim 0.1 \ \micron$.
Beneath the irradiated surface layer, dust particles experience a
head-wind, which drives them inward. 
For the specific case of a minimum-mass-solar-nebula, less than a
thousandth of the dust mass experiences outward flow.
If the stellar luminosity is 15 times brighter than the sun, however,
or if the gas disk mass is as small as $\sim 100 M_{\earth}$, then the
surface outflow can dominate the inward flux in certain radial
ranges, leading to the formation of rings or gaps in the dust
disks.
\end{abstract}



\section{Introduction}

Circumstellar disks around young stars are composed of gas and dust
particles, and in general, the dust and gas accretion velocities are not
equal. 
The gas suffers turbulent viscosity and accretes onto the central star
on a viscous time-scale of order $10^6 - 10^7$ yr (Hartmann et al. 1998).
The orbital evolution of dust particles, on the other hand, is
controlled by gas drag (Adachi, Hayashi, \& Nakazawa 1976;
Weidenschilling 1977).

Several factors affect the radial motion of particles.
First, gas drag in the radial direction entrains particles in the
accretion flow.
When the gas density is high or when the particle size is small, the
particles follow the gas motion, and accrete onto the star on the
viscous time-scale.
Second, angular momentum is transfered between particles and the gas.
The radial gas pressure gradient shifts the angular velocity of the gas
slightly from the Keplerian value.
Moreover, if starlight can irradiate particles, then radiation pressure
causes non-Keplerian deviations in the particle velocity as well.
Dust particles can thus gain or lose angular momentum through gas
drag and thus move outward or inward, depending on whether they
experience a net head-wind or a net tail-wind.

In a previous paper (Takeuchi \& Lin 2002, hereafter Paper I), we
studied particle motion in optically thick disks, using the assumption
that radiation pressure on the dust is negligible.
We found that dust particles smaller than $10 \ \micron$ accrete
with smaller velocity than the gas.
As the gas accretion proceeds, small dust particles are left behind in
the disk and the dust-to-gas ratio gradually increases.
Because circumstellar disks associated with young stars are generally
optically thick, the majority of dust particles are not directly
exposed to the stellar radiation, and the approximation used in Paper I
is applicable.
There are always, however, some particles near the disk surfaces that
are directly irradiated, and the motion of such particles are affected
by radiation pressure. 
Klahr \& Lin (2001) and Takeuchi \& Artymowicz (2001) studied the motion
of particles in optically thin gas disks around young Vega-type stars,
where the dust experiences both gas drag and radiation pressure.
Because the radiation pressure reduces the angular velocity of the
dust below that of the gas, the dust tends to move outward.
Thus, in a more refined description of an optically thick disk, one
expects that particles in the irradiated surface layer move
outward, while beneath the surface layer particles move inward.
If the surface outflow dominates the midplane inflow, then there is 
a net radial expansion that may induce the formation of rings or gaps.

In this paper, we study the surface dust outflow in optically thick disks due
to radiation pressure.
The mass flux of the surface outflow is calculated, which is compared to
that of the inflow inside the disk.
We find that the surface outflow is negligible in a
minimum-mass-solar-nebula model,
but that the surface outflow can dominate inflow at some disk radii in
some cases, including situations involving high stellar luminosity or
low gas mass. 
In such disks, the appearance of rings or gaps is expected.


\section{The Distribution of the Gas and the Dust \label{sec:distri}}

\subsection{The Gas Disk}

We consider circumstellar disks around young pre-main-sequence stars.
The disks are turbulent and are accreting onto the central stars, but
they are assumed to be passive, that is, the disks are primarily heated
by stellar radiation only.
When turbulent fluctuations are averaged out, the mean gas motion is
nearly in hydrostatic equilibrium and the radial accretion flow
is much slower than the azimuthal Kepler rotation.
In cylindrical coordinates ($r,\theta,z$), the balance of forces in
the $r$- and $z$-directions are
\begin{equation}
r \Omega_g^2 - \frac{G M r}{(r^2+z^2)^{3/2}} - \frac{1}{\rho_g}
\frac{\partial P_g}{\partial r} = 0 \ ,
\label{eq:gaseq_r}
\end{equation}
\begin{equation}
- \frac{G M z}{(r^2+z^2)^{3/2}} - \frac{1}{\rho_g} \frac{\partial
P_g}{\partial z} = 0 \ ,
\label{eq:gaseq_z}
\end{equation}
where $\Omega_g$ is the gas angular velocity, $M$ is the central star's
mass, $\rho_g$ is the gas density, and $P_g$ is the gas pressure.
For simplicity, we assume that the gas disk has power-law density and
temperature profiles in the radial direction, and is isothermal in the
vertical direction.
The temperature distribution is written as a power-law of index $q$,
\begin{equation}
T(r,z)=T_0 r_{\rm AU}^q \ ,
\end{equation}
where the subscript ``0'' denotes quantities at 1 AU, and a
non-dimensional quantity $r_{\rm AU}$ is the radius in AU.
The isothermal sound speed is $c=c_0 r_{\rm AU}^{q/2}$ and the gas
disk scale height is thus defined as
\begin{equation}
h_g(r) \equiv \frac{c}{\Omega_{\rm K,mid}} = h_0 r_{\rm AU}^{(q+3)/2} \ ,
\end{equation}
where $\Omega_{\rm K} = [GM/(r^2+z^2)^{3/2}]^{1/2}$ is the Keplerian
angular velocity and the subscript ``mid'' indicates the midplane value.
From the balance between the stellar gravity and the gas pressure
gradient in the $z$-direction (eq. [\ref{eq:gaseq_z}]), we have the
density distribution
\begin{equation}
\rho_g (r,z) = \rho_{0} r_{\rm AU}^p \exp \left( - \frac{z^2}{2
h_g^2} \right) \ ,
\label{eq:gasdensity}
\end{equation}
where a power-law in $r$ is assumed.
The surface density of the gas disk is given by
\begin{equation}
\Sigma_g (r) = \int_{-\infty}^{+\infty} \rho_g dz = \sqrt{2 \pi} \rho_0
h_0 r_{\rm AU}^{p_s} \ ,
\label{eq:gassurface}
\end{equation}
where $p_s = p+(q+3)/2$.
The angular velocity $\Omega_g$ of the gas is slightly different from the
Keplerian angular velocity $\Omega_{\rm K}$ because of the radial pressure
gradient in the gas.
From equation (\ref{eq:gaseq_r}), we have
\begin{equation}
\Omega_g^2 (r,z) = \left( 1 - \eta \right) \Omega_{\rm K}^2(r,z) \ ,
\label{eq:veldiff}
\end{equation}
where $\eta = -(r \Omega_{\rm K}^2 \rho_g)^{-1} \partial P_g/ \partial
r$ is the ratio of the gas pressure gradient to the stellar 
gravity in the radial direction.
As derived in equation (17) in Paper I, $\eta$ is written to order
$(h_g/r)^2$ as
\begin{equation}
\eta = -\left( \frac{h_g}{r} \right)^2 \left( p+ q+ \frac{q+3}{2}
\frac{z^2}{h_g^2} \right) \ .
\end{equation}
The disk has a turbulent viscosity $\nu$.
Turbulent disk viscosity has been ascribed to various mechanisms, including 
the convective instability (Lin \& Papaloizou 1980) and
the magneto-rotational instability (Balbus \& Hawley 1991).
In this paper, we simply model the viscous effect of turbulence using the
so-called $\alpha$ prescription (Shakura \& Sunyaev 1973),
\begin{equation}
\nu = \alpha c h_g = \alpha c_0 h_0 r_{\rm AU}^{q+3/2} \ .
\label{eq:alphavis}
\end{equation}
We adopt the following fiducial parameters:
$M = 1 \ M_\sun$, $\rho_0 = 2.83 \times 10^{-10}  \ {\rm g \ cm}^{-3}$,
$h_0 = 3.33 \times 10^{-2}$ AU, $p=-2.25$, $q=-0.5$, and
$\alpha=10^{-3}$. 
These values correspond to a disk having a gas mass of 
$2.5 \times 10^{-2} M_{\sun}$ inside 100 AU. The power law index of the
surface density is $p_s=p+(q+3)/2=-1.0$.
The temperature distribution is $T= 278 \ r_{\rm AU}^{-1/2} \ {\rm K}$.

\subsection{The Dust Disk}

\subsubsection{Radial Distribution}

The radial surface density distribution of the dust is assumed to
be proportional to that of the gas disk, so that $\Sigma_{d, {\rm all}}
(r) = f_{\rm dust} \Sigma_g (r)$, where the constant$f_{\rm dust}$ is the
dust-to-gas mass ratio. 
Throughout the paper, we adopt the value $f_{\rm dust} = 0.01$ at all radii.
The subscript ``all'' represents contributions from the entire
distribution of particle sizes.
In this paper, we consider cases where all particles have the same size
and also power-law size distributions.
For power-law size distributions, the size range is bracketed at the
large and small ends, $[s_{\rm min}, s_{\rm max}]$.
We adopt a power-law index $-3.5$, appropriate to the distribution
expected following collisional evolution (Hellyer 1970).
The surface number density of particles in a size range $[s,
s+ds]$ at a distance $r$ from the star is given by
\begin{equation}
N_{\Sigma} (r,s) ds = N_1 (r) s_{\rm cm}^{-3.5} ds \ ,
\label{eq:numdistr}
\end{equation}
where $s_{\rm cm}$ is the particle size in centimeters.
The radial distribution of particles $N_1$ is
\begin{equation}
N_1 (r) = \frac{3}{8 \pi \rho_p} f_{\rm dust} \Sigma_g(r) \left(
s_{\rm max,cm}^{1/2} -  s_{\rm min,cm}^{1/2} \right)^{-1} ({\rm
cm})^{-4} \ ,
\end{equation}
where $\rho_p$ is the physical density of a particle.
We adopt $\rho_p=1.25 \ {\rm g \
cm}^{-3}$, $f_{\rm dust} = 0.01$, $s_{\rm min} = 0.01 \ \micron$, and
$s_{\rm max} = 1 \ \micron$.

\subsubsection{Vertical Distribution}

Dust particles feel the $z$-component of the stellar gravity, and thus
sediment toward the disk midplane.
Small particles ($s \la 1$ cm) immediately (within one orbit)
reach the terminal velocity 
$v_{z,d}$, where the gravity and the gas drag are in balance.
For small particles considered here ($s \la 1$ cm), the Epstein's drag law
is applied, because the mean free path of the gas molecules (7 cm at 1
AU) is larger than the particle size.
The gas drag force is
\begin{equation}
{\mbox{\boldmath $F$}}_g = - \frac{4}{3} \pi \rho_g s^2  v_t
({\mbox{\boldmath $v$}}_d - {\mbox{\boldmath $v$}}_g ) \ ,
\end{equation}
where $v_t$ is the mean thermal velocity of the gas molecules, and
$\mbox{\boldmath $v$}_d$ and $\mbox{\boldmath $v$}_g$ are
the velocities of the dust and the gas, respectively.
The stopping time due to the gas drag is $t_s = m_d | {\mbox{\boldmath
$v$}}_d - {\mbox{\boldmath $v$}}_g | / |{\mbox{\boldmath $F$}}_g|$,
where $m_d = \case{4}{3}\pi \rho_p s^3$ is the particle mass.
We define the non-dimensional stopping time normalized by the orbital
time as,
\begin{equation}
T_s = t_s \Omega_{\rm K,mid}= \frac{\rho_p s v_{\rm K,mid}}{\rho_g r v_t} \ ,
\label{eq:stoptime}
\end{equation}
where $v_{\rm K, mid}$ is the Keplerian velocity at the midplane.
When terminal velocity is achieved, the gas drag balances the
$z$-component 
of the stellar gravity, $-m_d \Omega_{\rm K}^2 z$. Using the
approximation $\Omega_{\rm K} \approx \Omega_{\rm K, mid}$, the terminal
velocity $v_{z,d}$ is 
\begin{equation}
v_{z,d} = - \Omega_{\rm K, mid} T_s z \ .
\end{equation}
The mass flux of particles in the vertical direction is 
\begin{equation}
f_{z,{\rm sed}} = \rho_d v_{z,d} \ ,
\end{equation}
where $\rho_d$ is the dust mass density.
If the gas disk were laminar, the particles would concentrate at the
midplane in a time that is short compared to the evolution time
of the gas disk (the sedimentation time-scale is about $10^6$ yr
for $1 \ \micron$ particles at 10 AU).
When the gas disk is turbulent, however, the gas stirs and disperses the
particles and prevents sedimentation.
The mass flux of dust arising from this turbulent diffusion is assumed
proportional to the gradient of the dust concentration $\rho_d /
\rho_g$ in analogy with molecular diffusion (e.g, Monin \& Yaglom,
1971, p. 579; Morfill, 1985, \S3.5.1; Cuzzi, Dobrovolskis, \& Champney
1993).
In the vertical direction, the mass flux $f_{z,{\rm dif}}$ is
\begin{equation}
f_{z,{\rm dif}} = - \frac{\rho_g \nu}{\rm Sc} \frac{\partial}{\partial
z} \left( \frac{\rho_d}{\rho_g} \right) \ ,
\end{equation}
where the Schmidt number, ${\rm Sc}$, represents the coupling strength
between the particles and the gas. 
For small particles that are well coupled to the gas, ${\rm Sc}$
approaches unity.
For large particles that are decoupled from the gas, Sc becomes infinite. 
In our standard model, we use ${\rm Sc}=1$.

In equilibrium, the sedimentation mass flux balances the diffusive
mass flux:
\begin{equation}
f_{z,{\rm sed}} + f_{z,{\rm dif}} = 0 \ .
\end{equation}
The equilibrium dust density distribution is given by equation (31) in
Paper I.
For particles in a size range $[s, s+ds]$, the density distribution is
\begin{eqnarray}
\rho_{d}(r,z,s) ds & = & \rho_{d}(r,0,s) \exp \left[ 
- \frac{z^2}{2h_g^2} \right. \nonumber \\ 
& - & \left. \frac{{\rm Sc} T_{s, {\rm mid}}}{\alpha}
\left( \exp \frac{z^2}{2 h_g^2} - 1 \right) \right] ds \ , \nonumber \\
\label{eq:dustdistr}
\end{eqnarray}
where $T_{s, {\rm mid}}$ is the midplane value of the
non-dimensional stopping time.
The total dust density is 
\begin{equation}
\rho_{d,{\rm all}}(r,z) = \int_{s_{\rm min}}^{s_{\rm max}}
\rho_{d}(r,z,s) ds \ .
\end{equation}

As discussed in Paper I, the sedimentation of particles is more
effective for larger particles and at larger distances from the star.
As increasing the distance, the dust disk becomes relatively thinner
compared to the gas disk.

\subsection{Optical Depth from the Star}

The disks considered in this paper are optically thick at visible
wavelengths even in the vertical direction ($\tau_{\rm ver} \sim 100$ at
100 AU). 
Stellar light cannot penetrate the dust disk.
Only particles residing in the thin surface layer are exposed to
stellar radiation pressure.

The optical depth from the star to the position $(r,z)$ in the dust disk
is
\begin{equation}
\tau (r,z) = \int_{r_{\rm in}}^{r} \frac{\kappa}{f_{\rm dust}}
\rho_{d,{\rm all}}(r',z')
\sqrt{1 + \frac{z^2}{r^2} } dr' \ ,
\label{eq:optdepth}
\end{equation}
where $z' = (z / r)r'$, $r_{\rm in}$ is inner edge of
the dust disk, and $\kappa$ is the dust opacity in the visible
wavelength for an unit mass of the gas.
The opacity for a unit mass of the dust is $\kappa / f_{\rm dust}$.
In our calculations, we adopt $r_{\rm in}=0.1 \ {\rm AU}$.
This choice of inner boundary does not affect the integral
(\ref{eq:optdepth}), because the dust disk is flared and the main contribution
to the optical depth comes from the dust at relatively large distances from
the star.
We adopt a dust mass opacity $\kappa = 100 \ {\rm cm}^2 \ {\rm g}^{-1}$ for
visible wavelengths.
For particles smaller than $100 \ \micron$, the opacity is independent
of (the maximum) particle size (see Figs. 5 and 11 in Miyake \& Nakagawa
1993). 
Thus, we need not consider variations in $\kappa$ with changes in
the particle size.

\begin{figure}
\epsscale{1.0}
\plotone{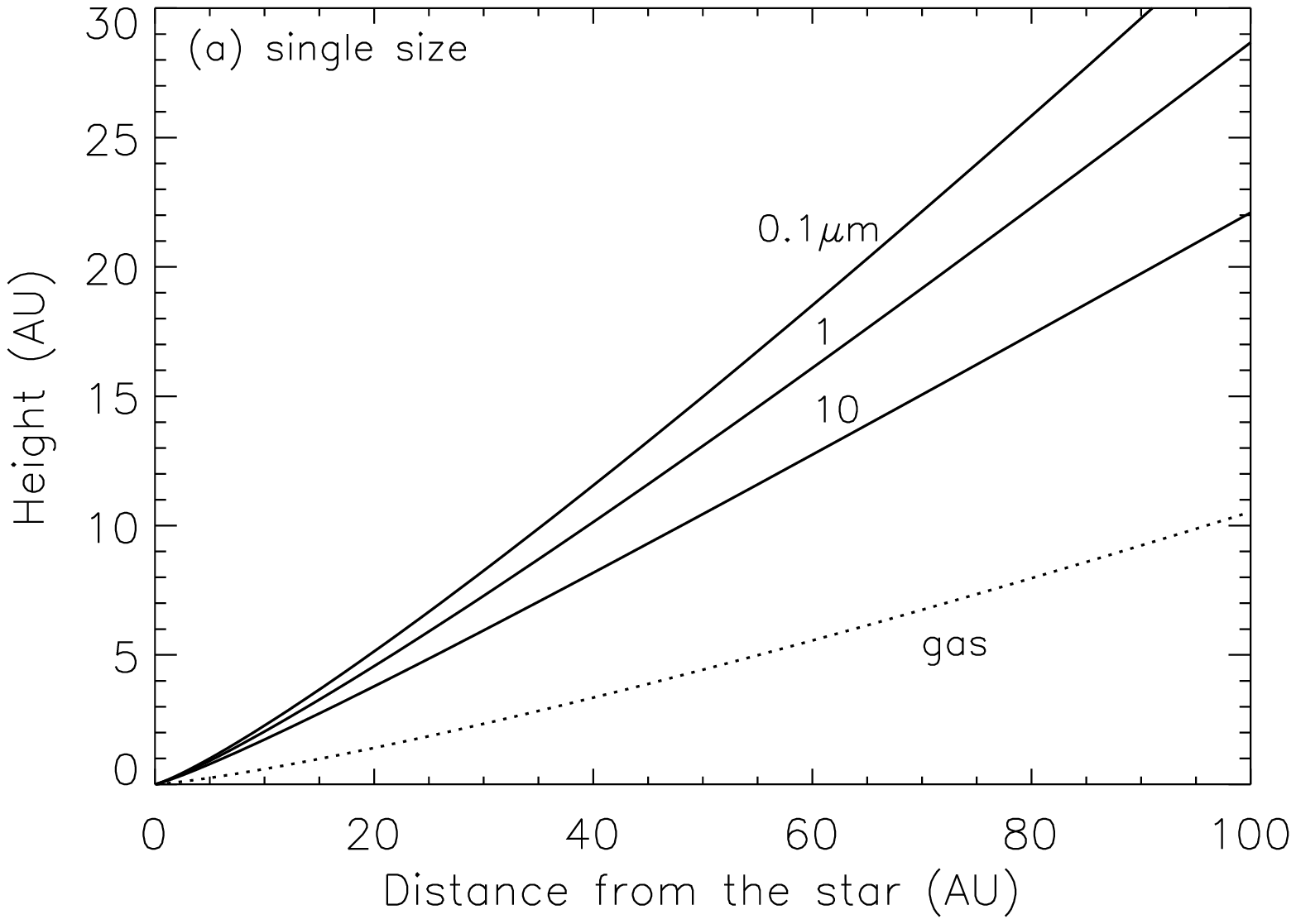}
\plotone{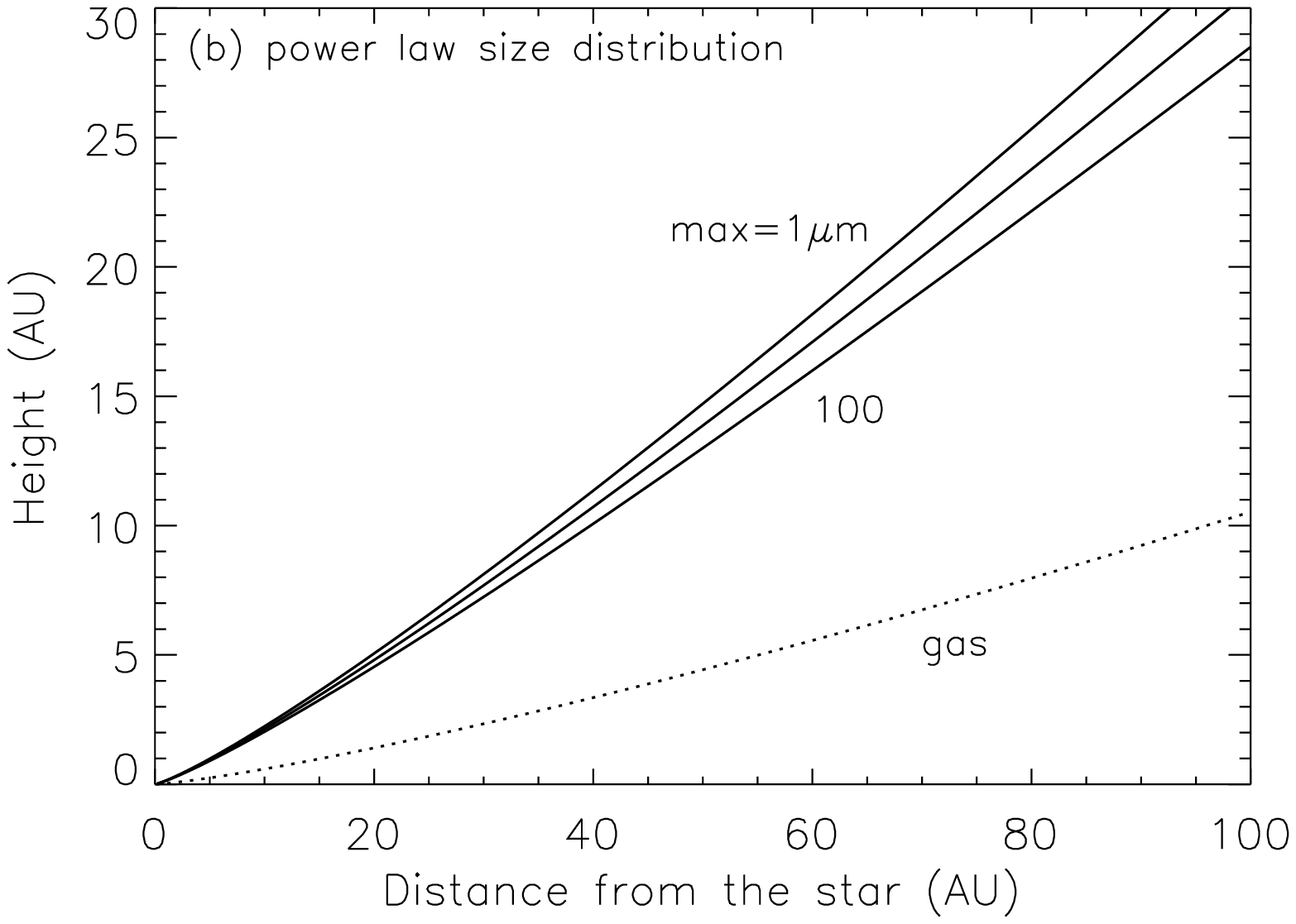}
\caption{
Locations of the illuminated surfaces.
($a$) The dust particles are assumed to be single-sized.
The sizes are $0.1$, $1$, and $10 \ \micron$ from the upper line.
The dotted line shows the scale height of the gas disk.
($b$) The size distribution is a power-law with an index $-3.5$.
The minimum size is $0.01 \ \micron$. The maximum sizes are $1$, $10$, and
$100 \ \micron$ from the upper line.
\label{fig:lightsur}
}
\end{figure}

Figure \ref{fig:lightsur} shows the disk surfaces at which the optical
depth to the starlight is unity.
Above these illuminated surfaces, the dust particles are directly
exposed to the stellar radiation.
Figure \ref{fig:lightsur}$a$ shows the illuminated surfaces of the standard
disk models which are composed of single-sized particles ($0.1$ to $10 \
\micron$).
As the particle size increases, the level of the illuminated surfaces
becomes lower.
This size dependence is because larger particles sediment more readily,
and thus
concentrate near the midplane.
The illuminated surfaces have a slightly flatter geometry than the gas disk,
because the dust particles at larger distances sediment more effectively.
Figure \ref{fig:lightsur}$b$ shows the illuminated surfaces of the models
containing power-law size distributions.
The illuminated surfaces lie at roughly three disk scale heights, 
which are similar to those of the single-sized
$0.1 \ \micron$ particles.
The location of the illuminated surface does not change significantly
with the maximum size of the particles, because high altitude above the
midplane is mostly populated by the smallest particles.


\section{Outflow in the Surface Layer}

Dust particles above the illuminated surface suffer radiation pressure
from the central star.
The radiation pressure balances a part of the stellar
gravity, so that the small particles orbit slower than the gas.
Because the gas drag transfers angular momentum from the gas to the
particles, the particles move outward, leading to an outflow in the
surface layers of dust disks.
In this section, we estimate the mass fluxes carried by these outflows.

\subsection{Outflow Mass Flux}

\begin{figure}
\epsscale{1.0}
\plotone{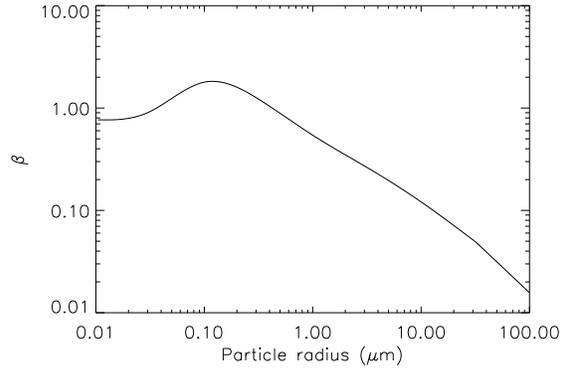}
\caption{
Radiation pressure to gravity ratio $\beta$ of the ``young cometary
particles.'' The value is from Wilck \& Mann (1996).
\label{fig:beta}
}
\end{figure}

The radiation pressure is expressed through the ratio $\beta$ to the
stellar gravity.
The radial component is 
\begin{equation}
F_{\rm rad} = \beta m_d r \Omega_{\rm K}^2 \ .
\end{equation}
The value of $\beta$, which is calculated with the Mie's scattering
theory, depends on the particle composition, porosity, shape, and size.
In this paper, we adopt $\beta$ for ``young cometary
particles,'' which is calculated by Wilck \& Mann (1996) for typical
interplanetary particles in the solar system and shown in
Figure \ref{fig:beta}.
In general, stellar light causes Poynting-Robertson drag on the
particles in addition to the gas drag induced by radiation pressure.
However, in the gas disks considered here, the gas density is
sufficiently large so that gas drag dominates Poynting-Robertson
drag (see the discussion in Takeuchi \& Artymowicz 2001, \S4.2).
Poynting-Robertson drag can therefore be neglected.

The equations of motion of a particle are
\begin{equation}
\frac{d v_{r,d}}{dt} = \frac{v_{\theta,d}^2}{r} - (1-\beta) \Omega_{\rm
K}^2 r - \frac{\Omega_{\rm K,mid}}{T_s} ( v_{r,d} - v_{r,g} ) \ ,
\label{eq:motion_r}
\end{equation}
\begin{equation}
\frac{d}{dt} \left( r v_{\theta,d} \right) =
 - \frac{v_{\rm K,mid}}{T_s} ( v_{\theta,d} - v_{\theta, g} ) \ ,
\label{eq:motion_th}
\end{equation}
where $v_r$ and $v_{\theta}$ are the $r$- and $\theta$- components of
the velocity, and the subscripts ``$g$'' and ``$d$'' distinguish gas and
dust. 
As in Paper I, we assume $v_{\theta,g} \approx v_{\theta,d} \approx
v_{\rm K,mid}$ and 
$d (r v_{\theta,d} ) / dt \approx v_{r,d} d (r v_{\rm K,mid} )/dr =
v_{r,d} v_{\rm K,mid}/2$.
From equation (\ref{eq:motion_th}) we then have
\begin{equation}
v_{\theta,d} - v_{\theta,g} = -\frac{1}{2} T_s v_{r,d} \ .
\label{eq:dvth}
\end{equation}
In equation (\ref{eq:motion_r}), $d v_{r,d} / dt = O(v_{r,d}^2/r)$ is
neglected if $(v_{r,d} / v_{\rm K,mid})^2 \ll 1$.
We also neglect second order terms in $\eta$ and $(v_{\theta,d} -
v_{\theta,g})/v_{\rm K,mid}$.
To this level of approximation, $v_{\theta, d}^2 \approx v_{\theta,g}^2
+ 2 v_{\theta,g}
(v_{\theta,d} - v_{\theta,g}) \approx v_{\rm K}^2 (1 - \eta) + 2 v_{\rm
K, mid} (v_{\theta,d} - v_{\theta,g})$, and 
equation (\ref{eq:motion_r}) becomes
\begin{equation}
(\beta - \eta) v_{\rm K}^2 + 2 v_{\rm K,mid} ( v_{\theta,d} -
v_{\theta,g} ) - \frac{v_{\rm K,mid}}{T_s} (v_{r,d}-v_{r,g}) = 0 \ .
\end{equation}
Using equation (\ref{eq:dvth}) and $v_{\rm K} \approx v_{\rm K,mid}$,
the particle radial velocity is 
\begin{equation}
v_{r,d} (r,z,s) = \frac{T_s^{-1} v_{r,g} + (\beta-\eta) v_{\rm
K,mid}}{T_{s} + T_{s}^{-1}} \ .
\label{eq:vrout}
\end{equation}
The value of $\eta$ is of order $(h_g/r)^2$ and it is smaller than $0.05$.
Since particles smaller than $10 \ \micron$ have $\beta \gg \eta$ (see
Fig. \ref{fig:beta}), $\eta$ in equation (\ref{eq:vrout}) can be neglected.
Particles larger than $\sim 100 \ \micron$ may have $\beta$ smaller than
$\eta$.
For such large particles, the radiation pressure is unimportant, even if
they are above the illuminated surface. In this section, we
consider only small particles with radii less than $10 \ \micron$.
Such particles have non-dimensional stopping times $T_s \ll 1$ (see
Fig. 2 in Paper I). 
The radial velocity then reduces to 
\begin{equation}
v_{r,d} = v_{r,g} + \beta T_s v_{\rm K,mid} \ .
\label{eq:vrout2}
\end{equation}
The mass flux of the outflow in the surface layer is
\begin{equation}
F_{\rm sur} (r) = \int_{s_{\rm min}}^{s_{\rm max}} 2 \int_{z_{\rm
sur}}^{\infty} 2 \pi r \rho_d (r,z,s) v_{r,d} (r,z,s) dz ds \ ,
\end{equation}
where $z_{\rm sur}$ is the height of the illuminated surface.
The factor 2 arises from considering both the top and bottom surfaces.
In the integration in $z$ above the illuminated surface, $v_{r,d}$
varies less rapidly than $\rho_d$, and we may use its value at the
illuminated surface.
The mass flux is then expressed as
\begin{equation}
F_{\rm sur} (r) =  2 \pi r \int_{s_{\rm min}}^{s_{\rm max}} v_{r,d} (r,z_{\rm
sur}, s) \Sigma_{d, {\rm sur}} (r, s) ds \ , 
\end{equation}
where $\Sigma_{d, {\rm sur}} ds$ is the surface density of dust
particles above the illuminated surface in the size range $[s, s+ds]$,
\begin{equation}
\Sigma_{d, {\rm sur}} (r, s) =  2 \int_{z_{\rm
sur}}^{\infty} \rho_d (r,z,s) dz \ .
\end{equation}
The surface outflow can clear out the dust within a distance $r$ from the
star on a time-scale $t_{\rm out} \approx M_{\rm dust}(r) / F_{\rm
sur}(r)$, where $M_{\rm dust}(r)$ is the dust mass within $r$.
In our models, the disk mass is dominated by the outermost part of the
disk and $M_{\rm dust}(r) \approx \pi r^2 \Sigma_{d, \rm all} (r)$. 
The dust clearing (or the disk evolution) time-scale is thus
\begin{equation}
t_{{\rm out}} (r) \approx \frac{\pi r^2 \Sigma_{d, {\rm all}}(r)}{F_{\rm
sur}(r)} \ .
\label{eq:tout}
\end{equation}

While at the surface layer particles flow outward, the particles beneath
the surface layer are shielded from stellar irradiation.
At such locations, the gas drag causes a net inward flow of particles as
shown in Paper I. 
From equation (\ref{eq:vrout}) with $\beta=0$, the radial velocity of
particles beneath the surface layer is
\begin{equation}
v_{r,d} (r,z,s) = \frac{T_s^{-1} v_{r,g} - \eta v_{\rm
K,mid}}{T_{s} + T_{s}^{-1}} \ .
\label{eq:vrin}
\end{equation}
The mass flux is
\begin{equation}
F_{\rm in} = \int_{s_{\rm min}}^{s_{\rm max}} \int_{-z_{\rm
sur}}^{z_{\rm sur}} 2 \pi r \rho_d (r,z,s) v_{r,d} (r,z,s) dz ds \ .
\end{equation}
The time required for the inflow to transport a mass equal to the dust
mass within $r$ is
\begin{equation}
t_{{\rm in}}(r) \approx \frac{\pi r^2 \Sigma_{d, {\rm all}}(r)}{| F_{\rm
in}(r) |} \ .
\end{equation}
We define this as the disk evolution time through the inflow.
This definition of $t_{\rm in}$ does not mean the time for carrying the
total dust mass outside $r$ to the inner part, but it is convenient for
comparing with $t_{\rm out}$.
Comparison of $t_{\rm out}$ and $t_{\rm in}$ simply means comparison of
the outflow and inflow fluxes.
The disk clearing through the surface outflow occurs effectively only if
the surface flux is stronger than the inflow flux, i.e., $t_{{\rm out}} <
t_{{\rm in}}$.
The comparison of these two time-scales is discussed in \S\ref{sec:inflow}
 and \S\ref{sec:models} below.

\subsection{Evolution Time of the Disks through the Outflow}

We first discuss the disk clearing time, $t_{\rm out}$, through the
surface outflow.
In this subsection, we adopt the assumption that inflow flux near the
midplane is negligible and that only the surface flow affects the dust
structure within the disk.
In \S\ref{sec:inflow} and \S\ref{sec:models}, we relax this assumption and
consider more realistic scenarios including the effects of inflow near
the midplane.

\begin{figure}
\epsscale{1.0}
\plotone{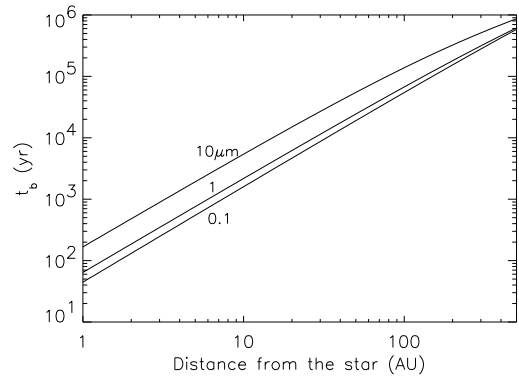}
\caption{
Blow-off time of the outflowing particles in the surface
layer. The particles are single-sized of $10$, $1$, and $0.1 \ \micron$ from
the upper line.
\label{fig:tb}
}
\end{figure}

The particles in the surface layer have outflow velocity, $v_{r,d}
(r,z_{\rm sur}, s)$.
These particles are blown from their orbits to larger radii on a
time-scale $t_b = r/v_{r,d} (r,z_{\rm sur}, s)$.
For disks composed of single-sized particles, $t_b$ is plotted in
Figure \ref{fig:tb}.
The blow-off time for $0.1$ to $10 \ \micron$ particles is about $10^2$
yr at a few AU from the star, and is about $10^5$ yr at 100 AU.
The blow-off time is much shorter than the inward migration time inside the
disk, which is discussed in \S\ref{sec:inflow} below, because $\beta \gg
\eta$ and because the stopping time $T_s$ at the illuminated surface is
much longer than at the midplane.

\begin{figure}
\epsscale{1.0}
\plotone{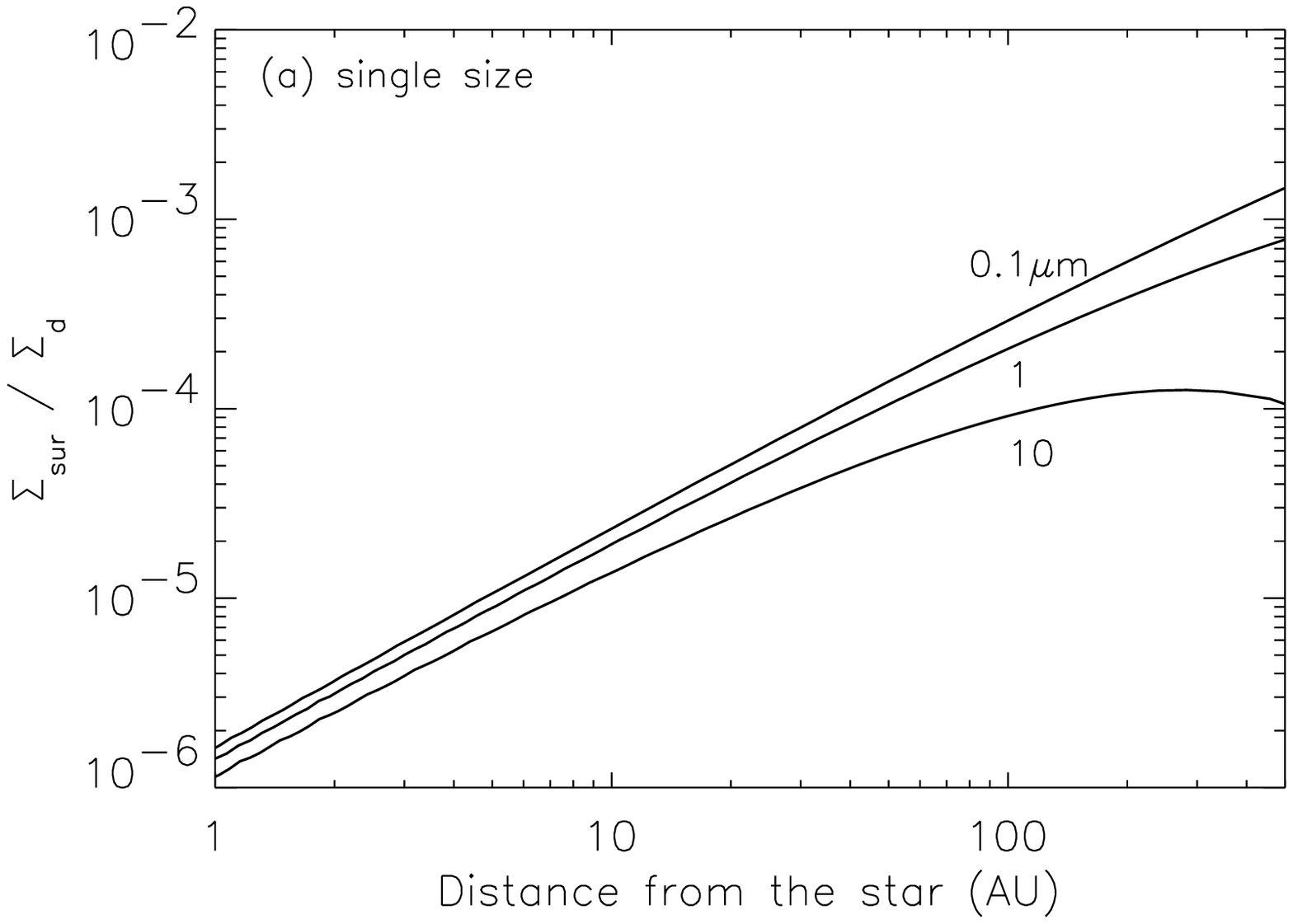}
\plotone{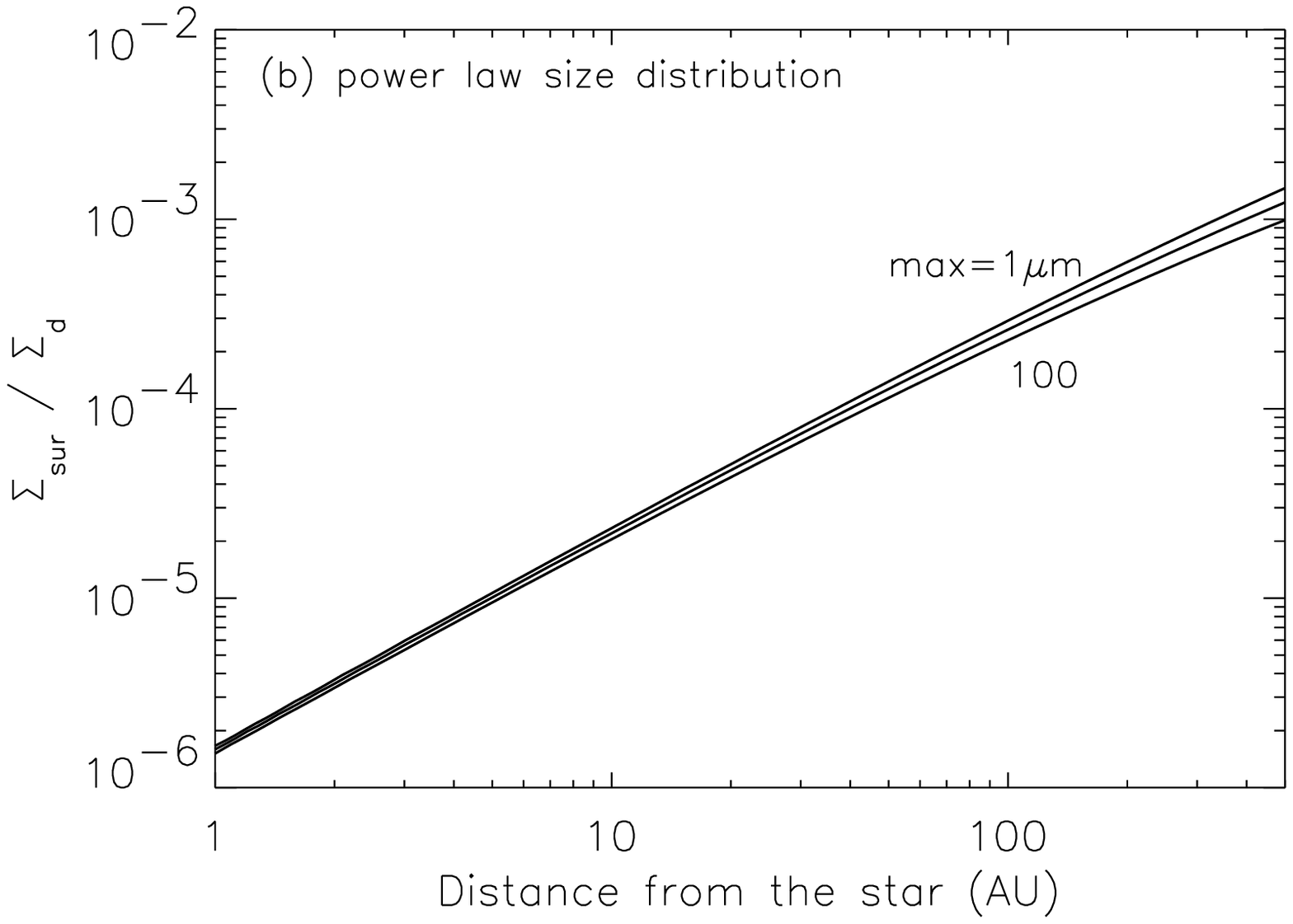}
\caption{
Mass ratio of the surface layer above the illuminated surface to the whole dust
disk ($\Sigma_{d, {\rm sur,all}} / \Sigma_{d, {\rm all}}$).
($a$) The dust particles are single-sized. The sizes are $0.1$, $1$, and
$10 \ \micron$ from the upper line.
($b$) The size distribution is a power-law with an index $-3.5$.
The minimum size is $0.01 \ \micron$. The maximum sizes are $1$, $10$, and
$100 \ \micron$ from the upper line.
\label{fig:surmass}
}
\end{figure}

In our standard model, the dust disk is highly optically thick.
Even in the vertical direction at 100AU, the optical depth is $\tau_{\rm
ver} \sim 100$, so the layer above the illuminated surface is very thin.
Figure \ref{fig:surmass} shows the ratio of the surface density of the
layer to that of the whole disk ($\Sigma_{d, {\rm sur,all}} / \Sigma_{d,
{\rm all}}$).
As the distance from the star increases, the mass ratio increases and the
surface layer becomes thicker.
However, even at 100 AU, the mass of the surface layer is less than
$10^{-3}$ times the dust disk mass.
In the case of single-sized particles (Fig. \ref{fig:surmass}$a$) the
mass of the surface layer is smaller for disks with larger particle 
sizes.
As the particles grow larger, they sediment more to the midplane, and the
surface layer becomes thinner.
On the other hand, for the mixed-sized particles
(Fig. \ref{fig:surmass}$b$), the mass of the surface layer 
does not vary significantly with the maximum size.
Because smaller particles exist at higher altitudes above the midplane,
the surface layer is composed mainly of the smallest particles, and the
quantity of smallest particles in the surface layer does not vary
with the maximum size.

\begin{figure}
\epsscale{1.0}
\plotone{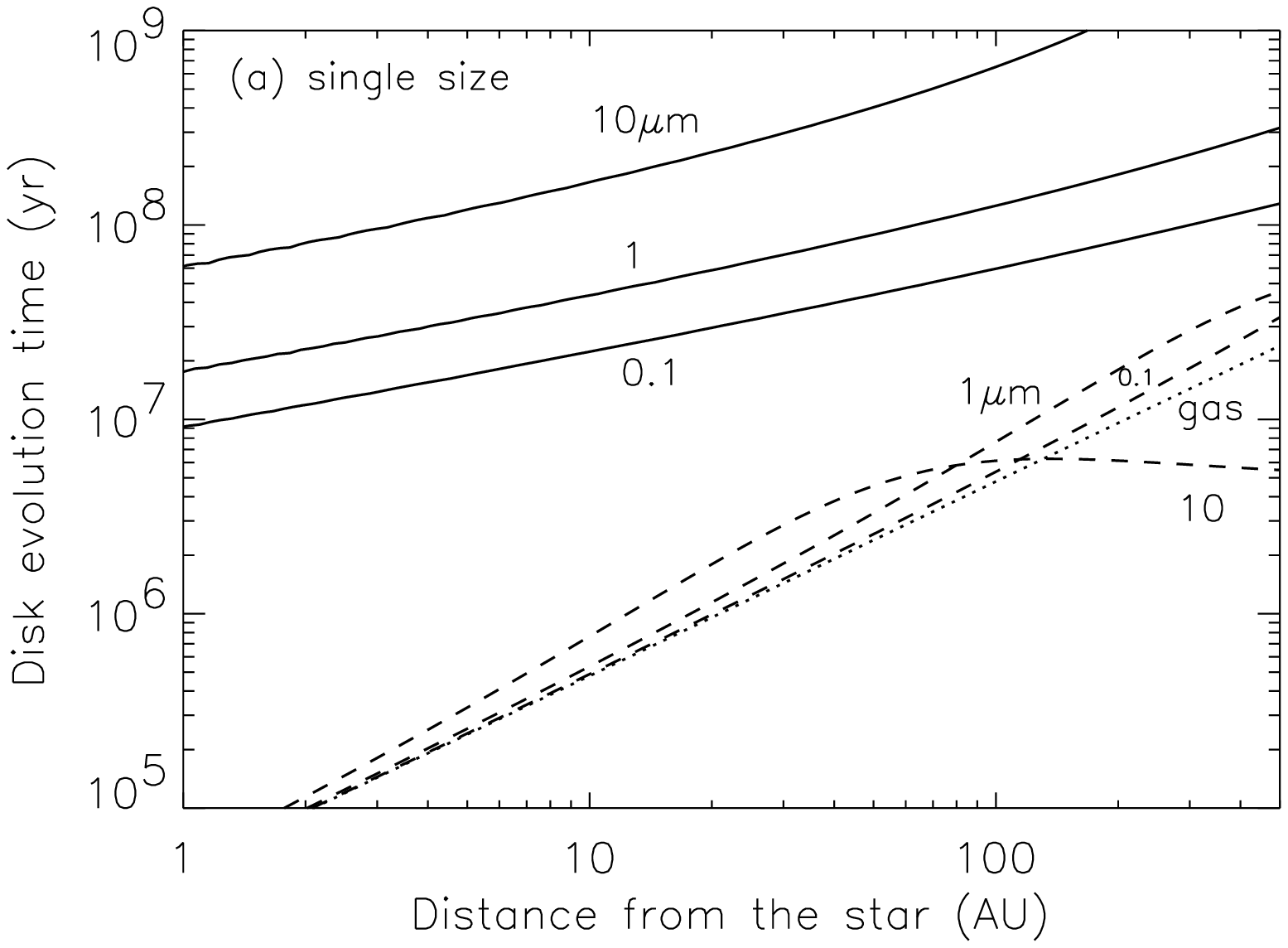}
\plotone{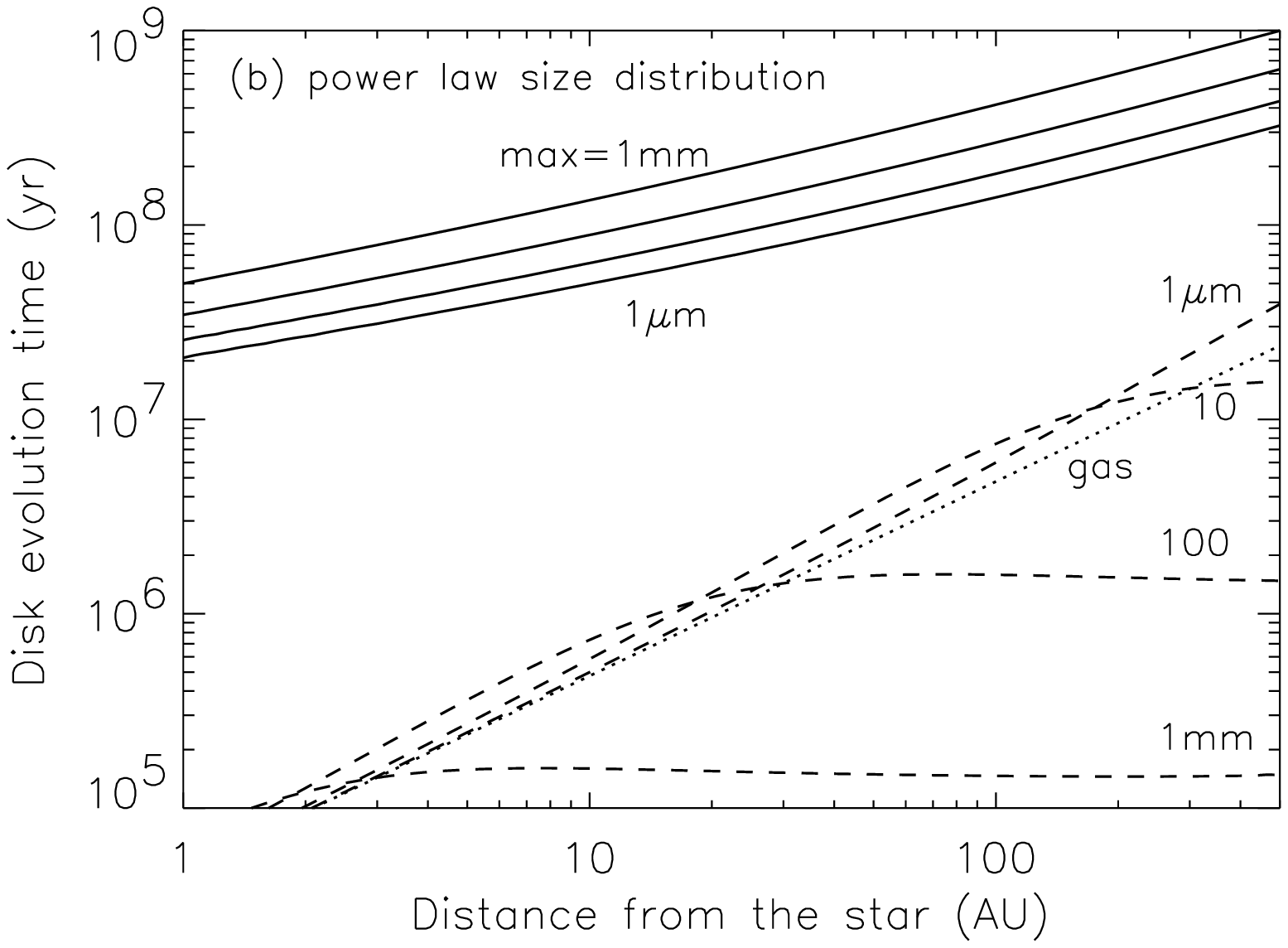}
\caption{
Disk evolution time through the outflow and the inflow. The solid lines
show the evolution time through the surface outflow, and the dashed
lines show the evolution time through the inflow inside the disk. 
The dotted line shows the accretion time-scale of the gas.
($a$) The particles are single-sized. The sizes are $10$, $1$, and $0.1 \
\micron$ from the upper lines (at 10 AU for the dashed lines). 
($b$) The size distribution is a power-law with an index $-3.5$ and a
minimum size  $0.01 \ \micron$. The maximum sizes are $1$ mm, $100$, $10$,
and $1 \ \micron$ from the upper solid line and from the lower dashed
line.
\label{fig:tout}
}
\end{figure}

Although Figure \ref{fig:tb} implies that the outflow velocity at the
surface layer is high, the outflow mass flux is not large because of the
small mass contained in the surface layer.
The disk evolution time $t_{\rm out}$ through the
surface outflow is shown in Figure \ref{fig:tout}$a$. 
The evolution time of the disk composed of $0.1 \ \micron$ particles is
about $10^7 \ {\rm yr}$ at 1 AU and $10^8 \ {\rm yr}$ at 100 AU.
Thus, in $10^7$ yr, the dust outflow mass which has passed 1 AU becomes
comparable to the dust disk mass inside 1 AU.
If the radial motion of the dust disk occurs only through surface
outflow, the particles in the innermost part of the disk will escape
from the region inside 1 AU in $10^7$ yr, and a hole will appear.
The hole expands, and in $10^8$ yr, the radius of the hole reaches 100
AU.
There is, however, an accretion flow under the surface layer.
Inhibition of disk clearing by the accretion flow is discussed in
\S\ref{sec:inflow} below.
As the particle size increases, the evolution time increases as well.
This is because particles with larger sizes concentrate more at the
midplane and the surface layer becomes thinner.
The blow-off time $t_b$ for a larger particle at its surface layer is also
larger because of the higher local density of the gas at the lower surface
layer (Fig. \ref{fig:tb}).
If the disk is composed of $10 \ \micron$ particles, it takes about
$10^9 \ {\rm yr}$ for a disk of radius 100 AU to evolve.

Figure \ref{fig:tout}$b$ shows the evolution times for disks composed
of mixed-sized particles.
The disk containing $[0.01 \ \micron-1 \ \micron]$ particles evolves in about
$3 \times 10^7$ yr at 1 AU and in $3 \times 10^8$ yr at 100 AU.
The disk evolution occurs mainly through the outflow of $0.1 \
\micron$ particles.
While the surface layer is composed primarily of the smallest ($0.01 \
\micron$) particles, these particles do not receive radiation pressure as
effectively  as particles with $0.1\ \micron$ (see Fig. \ref{fig:beta}).
Thus, the surface outflow is mainly carried by a minor fraction of
particles in the $0.1 \ \micron$ range, and the disk evolution time of
the mixed-sized 
particles is much longer than that of the disk with the single-sized
particles of $0.1 \ \micron$.
The evolution time increases with the maximum particle size.
The dependence of the maximum size on the evolution time, however, is not
as significant as in the case of single-sized particles.
For disks composed of mixed-sized particles, particles of size $0.1
\ \micron$ form the main contribution to the surface outflow.
The number density of $0.1 \ \micron$ particles decreases as the maximum
particle size increases.
However, because the thickness of the surface layer is determined by the
optical depth from the star and the opacity comes mainly from $0.1 \
\micron$ particles, the total mass of $0.1 \ \micron$ particles in the
surface layer does not change significantly (Fig. \ref{fig:surmass}$b$).
Hence, the outflow flux maintains a relatively constant value.

\subsection{Comparison with the Inward Migration Time \label{sec:inflow}}

The surface outflow leads to disk clearing only if it dominates the
under-surface inflow.
The disk evolution time-scale, $t_{\rm in}$, through the inflow is
shown as the dashed lines in Figure \ref{fig:tout}.
As discussed in Paper I, for sufficiently small particles that are well coupled
to the gas, the inflow velocity is similar to the accretion velocity
of the gas.
In the standard model, particles smaller than $1 \ \micron$ have
velocities nearly indistinguishable from the gas throughout the disk.
Particles of $10 \ \micron$ decouple from the gas motion at large
distances ($r \ga 50$ AU) and have a shorter evolution time than the gas
and smaller particles.
The inflowing mass flux is much larger (10 times or more) than the 
mass flux of the surface outflow even for $0.1 \ \micron$ particles.
Therefore, the net mass flux of particles is dominated by the inflow
beneath the surface layer.
In the disks composed of mixed-sized particles (Fig. \ref{fig:tout}$b$),
the inflow also dominates the net flow.
The surface outflow has a very minor effect on the disk evolution in our
standard model, regardless of the particle size.

\subsection{Various Models \label{sec:models}}

In our standard model, the surface outflow is negligible for disk
evolution compared to the inflow inside the disk.
This flow outcome is due to the surface layer being very thin such that
its mass is a small fraction of the entire dust disk mass.
The surface outflow may be important under some conditions where the
surface layer is thicker or the outflow is much faster than the standard
model.
In this subsection, we consider several models to see under which
conditions the surface outflow is important.

\subsubsection{High Luminosity \label{sec:highl}}

\begin{figure}
\epsscale{1.0}
\plotone{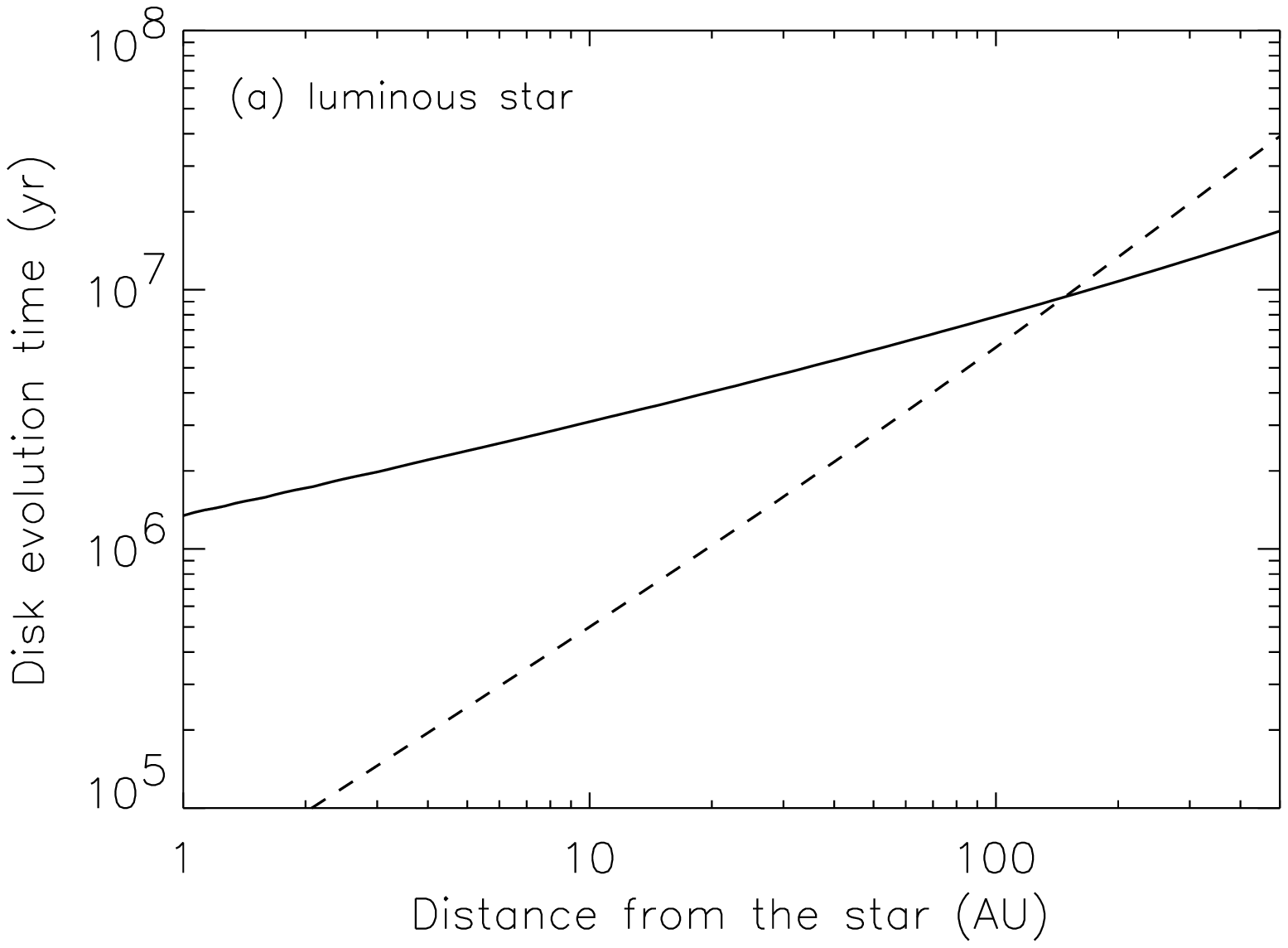}
\plotone{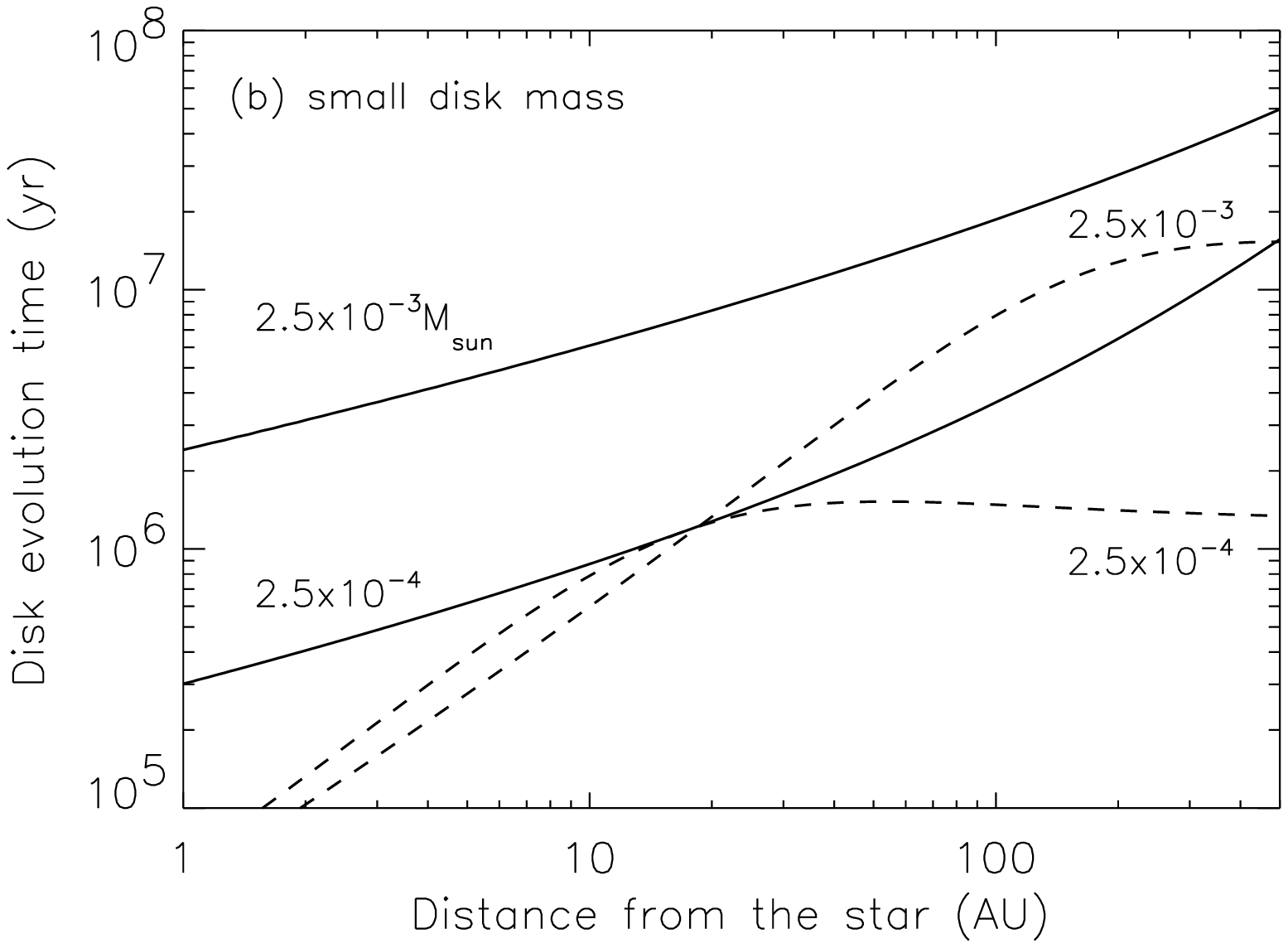}
\caption{
Disk evolution time for various models. The solid and dashed lines
correspond to the evolution time through the surface outflow and through
the inflow inside the disk, respectively.
Particles have a power-law size distribution with $s_{\rm max}=1 \
\micron$. Parameters are the same as the standard model except for:
($a$) The luminosity-mass ratio of the central star is $L/M = 15
L_{\sun}/M_{\sun}$. 
($b$) The disk is less massive. The gas masses inside 100 AU are $2.5
\times 10^{-3} M_{\sun}$ and $2.5 \times 10^{-4} M_{\sun} = 83
M_{\earth}$ from the upper line.
($c$) The particles is more diffusive (${\rm Sc} = 0.1$).
\label{fig:tout2}
}
\end{figure}

Massive young stars have a larger luminosity-mass ratio. 
For example, $2.5 M_{\sun}$ stars have about $20 L_{\sun}/M_{\sun}$ for
$\sim 10^7$ yr (D'Antona \& Mazzitelli 1994).
Under such a high luminosity-mass ratio, the dust particles are exposed
to a stronger radiation pressure ($\beta$ is proportional to $L/M$).
Figure \ref{fig:tout2}$a$ shows the evolution time of the disk in the
case where the stellar luminosity-mass ratio is 15 times larger than the
standard model.
The outflow velocity is 15 times faster than in the standard model, causing
the outflow mass flux to dominate the inflow at large distances ($r \ga
200$ AU).
In this model, dust particles which are initially outside $200$ AU are expelled
to larger distances, while particles inside $200$ AU are accreted onto the
star.
It is expected that a gap around $200$ AU forms in $10^7$ yr.

\subsubsection{Small Disk Mass}

If the total dust mass is smaller, the disk optical depth is smaller so that
the layer above the illuminated surface is thicker.
Figure \ref{fig:tout2}$b$ shows the evolution time of the disks whose
masses are $0.1$ and $0.01$ times smaller than the standard disk.
The mass flux of the surface outflow is almost independent of the disk
mass because the mass of the surface layer, which is determined by the
optical thickness, is approximately constant.
Therefore, the relative importance of the surface outflow, which is
proportional to the mass ratio of the surface layer to the whole disk,
increases with decreasing the total dust mass.
The evolution time through the inflow (see the dashed lines) at the
inner part of the disk is similar to that of the standard model, while
at the outer part it decreases with decreasing disk mass.
As the disk mass decreases, the gas density at the outer region of the
disk  becomes
insufficient to keep dust particles moving together with the gas.
The inward drift velocity of particles relative to the gas becomes
faster for disks with smaller gas mass.
For such drifting particles, the evolution time is proportional to
$r^{2+p-q/2}$ (see Paper I, \S3.3.3), which is constant with $r$ in our
models ($p=-2.25$ and $q=-0.5$).
For the disk with one tenth the mass as the standard model, the evolution time
through the surface outflow is still longer than that of the inflow.
If the disk is $10^{-2}$ times smaller than the standard model (the gas
mass inside 100 AU is 
$2.5 \times 10^{-4} M_{\sun} = 83 M_{\earth}$), the surface outflow flux
is comparable to the inflow flux at around 20 AU.
Thus, there is no net flow at radii in the neighborhood of 20 AU.
The dust particles initially outside 20 AU move inward until
their migration is terminated at 20 AU.
At the same time, the particles initially inside 20 AU continue to move
inward and accrete onto the star.
Therefore, the particle migration induces accumulation of dust at 20 AU
and forms a ring there.
Gas disks around young stars are expected to dissipate in $10^7$ yr
(Hartmann et al. 1998).
When the gas mass is reduced to about $100 M_{\earth}$, a ring
appears at a radius of a few tens of AU.

\subsubsection{High Diffusivity of Dust}

In the standard model, the diffusivity of dust particles subject to
turbulent motion of the gas is assumed to be the same as the diffusivity of
the gas, i.e., the Schmidt number, Sc, is unity.
Experiments show that Sc can be as small as 0.1 and
dust particles are much more diffusive than the gas, if the Stokes
number, ${\rm St}$, is of order $10^{-2}$ (e.g., Fig. 1 in Cuzzi et al.
1993).
The Stokes number is considered to be of the same order as the
non-dimensional stopping time $T_s$.
At radii of several hundred AU, $T_s \sim 10^{-2}$ for $0.1 \ \micron$
particles at the illuminated surface ($z \sim 3 h_g$).
Thus, it is expected that such particles will spread to higher altitude, where
they are easily blown off.
Figure \ref{fig:tout2}$c$ shows the evolution time for such particles with 
${\rm Sc} = 0.1$.
The dust particles are more diffusive and are stirred up to higher
altitude by turbulence than in the case of ${\rm Sc} = 1$. The illuminated
surface is located higher, where the gas density is lower and the outflow
velocity is faster.
The outflow mass flux, however, is still smaller than the inflow mass
flux within $r \la 500$ AU.
Inflow under the surface layer dominates even for the particles with
high diffusivity.

\subsubsection{Particle Growth \label{sec:growth}}

In the above discussion, we mainly used models which have a power-law size
distribution with a maximum size of $1 \ \micron$.
The dust particles in circumstellar disks may have evolved to sizes larger
than $1 \ \micron$ (Beckwith \& Sargent 1991; Mannings \& Emerson 1994;
D'Alessio, Calvet, \& Hartmann 2001).
Figure \ref{fig:tout}$b$ shows the evolution time of the disks as the
maximum size of the particles is varied.
The evolution time through the surface outflow does not change
significantly as discussed in \S3.1.
The evolution time through inflow inside the disk, however, rapidly
decreases with the growth of particles.
At the outer part of the disk where the gas density is insufficient
to keep the largest particles moving together with the gas, the
inflow velocity of the particles is proportional to their size.
For the size distribution with a power-law index $-3.5$, the dust mass
is dominated by the largest particles, and the inflow 
mass flux is proportional to the size of the largest particle.
The evolution time through the inflow is as small as $10^5$ yr when the
largest particle is $1$ mm, which is much shorter than that through the
surface outflow.
As the particles grow larger, the surface outflow becomes less important
for the overall disk evolution.

\section{Discussion and Summary}

\subsection{Progress in the Disk Evolution}

We compared the surface outflow flux to the under-surface inflow flux and
discussed conditions for disk clearing.
In those comparisons, we assumed the dust density profiles described in
\S\ref{sec:distri}, and neglected time variations of the dust density.
In real disks, however, the dust density varies as the surface outflow
proceeds.
We discuss how density changes affect the subsequent evolution.

As dust particles above the illuminated surface are blown off by the
stellar radiation, new particles continually diffuse into and replenish
this region from below.
The diffusion time over which the surface layer is re-populate is
\begin{equation}
t_{{\rm rp}} \approx \frac{{\rm Sc} (3 h_g)^2}{\nu} \approx \frac{9}{\alpha}
\Omega^{-1} \approx 10^3 r_{\rm AU}^{3/2} \ {\rm yr} \ ,
\end{equation}
where we use the fact that the distance between the surface layer and
the midplane is about three disk scale heights for $\alpha =
10^{-3}$ and ${\rm Sc}=1$.
The diffusion time ($10^6$ yr at 100 AU) is much shorter than the disk
evolution time $t_{\rm out}$. 
Thus during the disk evolution, the dust in the surface layer is
continuously replenished, and the surface outflow
continues until the dust in the midplane is depleted.

In disks composed of mixed-sized particles, small ($s \la 0.1 \
\micron$) particles constitute the majorities of the surface outflow.
The small particles are rapidly removed from the disk,
and are continuously replenished through the collisional destruction of
larger particles.
The collisional time scale, $t_{\rm col} = (\pi s^2 n_d v_{\rm
rel})^{-1}$, where $n_d$ is the particle number density and $v_{\rm
rel}$ is the relative velocity of particles, is much shorter than the
disk evolution time.
For example, if we assume that the majority of the dust is composed of 1
cm particles, the relative 
velocity is about $10 \ {\rm cm \ s}^{-1}$ at 1 AU (from Fig. 3 in
Weidenschilling \& Cuzzi 1993), and
the collisional time is $t_{\rm col} \sim 10^3$ yr. If the dust is 
composed of smaller particles, the collisional time is smaller.
The outcome of the collisions, however, is not well
understood, and we do not know how many small particles are produced
through the collisional destruction.
In this paper, we simply assume that the particle size distribution is in
an equilibrium state. The minimum size, the maximum size, and the power-law
index of the distribution do not vary with time.
In a realistic situation, however, at least the maximum size would grow
with time, leading to inhibition of the disk clearing, as discussed in
\S\ref{sec:growth}.
Calculations that take account of dust growth are needed to solve this issue.

Because the disk evolution time $t_{\rm out}$ is shorter at smaller
radii, the dust outflow proceeds from the inner disk.
The particles blown from the inner disk accumulate in the outer disk.
This process increases the dust-to-gas ratio at large radii.
We can ask whether this increase in the dust-to-gas ratio affects 
the subsequent disk evolution.
Suppose that the dust particles inside a radius $r$ are removed and
an inner hole (or a gap) is produced.
The removed particles pile up in an annulus just outside the
inner hole.
The width of the annulus is larger than the disk scale height. 
On the diffusion time $t_{\rm rp}$, the accumulated particles diffuse
over a few disk scale heights both in the vertical and in the radial
directions, and during a disk evolution time $t_{\rm out} > t_{\rm rp}$,
the particles diffuse even further.
Thus, the surface density of the annulus increases at most by a factor
$\sim \pi r^2 \Sigma_{d,\rm all} / (2 \pi r \Sigma_{d,\rm all} dr) \la
r/h_g \sim 10 $, where $dr \ga h_g \sim 0.1 r$ is the width of the
annulus.
Because the gas density does not change, the dust-to-gas ratio of the
annulus becomes $\la 10$ times larger.
The opacity $\kappa$, which is defined for a unit mass of the gas and is
proportional to the dust-to-gas ratio, also increases by a factor of
order $\la 10$ times.
(If the particles diffuse as far as the inner hole radius $r$ in a
disk evolution time $t_{\rm out}$, then the hole is refilled and dust
removal is inhibited. The time-scale for the refill is comparable to the
gas accretion time-scale $r^2/\nu$, which is shown as the dotted lines
in Figure \ref{fig:tout}.
In the standard model, the refill effectively inhibits the dust
removal. Only in the high stellar luminosity or the low mass disk cases,
the surface outflow can dominate the refill.)

Figure \ref{fig:flux} shows how the increase in the dust-to-gas ratio
affects the outflow mass flux.
In Figure \ref{fig:flux}, the outflow flux in the cases for $f_{\rm
dust}=0.1$ (i.e., the dust-to-gas ratio is 10 times larger than the
standard model) and for $f_{\rm dust}=10^{-3}$ (10 times smaller
dust-to-gas ratio) are shown with the standard model.
It is seen that the outflow flux depends only weakly on the dust-to-gas
ratio.
The change in the dust-to-gas ratio has a slight effect on the location
of the surface layer, but it hardly changes the dust mass of the surface layer.
The thickness of the surface layer is determined so as to make the
optical depth be unity, and its mass content 
does not depend on the dust-to-gas ratio.
The variation in the outflow flux arises mainly from change in the
location of the surface layer and from change in the 
angle of the incident starlight to the surface layer, and these effects
are small.
Therefore, the accumulation of blown off particles increases the
dust-to-gas ratio at the disk edge just outside the inner hole, but does
not significantly decrease the outflow flux there.
Even after the increase in the dust-to-gas ratio in the outer disk, the
surface outflow proceeds and the disk evolution continues with the 
time-scale plotted in Figures \ref{fig:tout} and \ref{fig:tout2}.

\subsection{Gap Formation in Herbig Ae/Be Disks}

The phenomenon of surface stripping can effectively evolve the disk
structure around young luminous stars whose luminosity-mass ratios are
larger than $\sim 15$ times the solar value, as shown in \S\ref{sec:highl}.
In the outer part ($\ga 200$ AU) of such disks, the surface outflow
dominates the inflow through the midplane, causing a net outward flow.
On the other hand, within 200 AU, the dust flows inward onto the central
star.
Around 200 AU, the dust density decreases.
The outflow or inflow flux around 200 AU is strong enough that it can
carry the total amount of the dust inside 200 AU in $10^7$ yr.
This flux maintains a relatively constant value throughout the evolution
even after the dust density and the dust-to-gas ratio have decreased (see
Fig. \ref{fig:flux}).
Consequently, the dust density declines at an almost constant rate.
The time-scale for dust diffusion to transport dust from the
midplane to replenish the surface layer is about $3 \times 10^6$ yr at 200 AU.
We expect the dust density at 200 AU will decline considerably in
$10^7$ yr.

The formation of gaps can be examined by radio observations.
At 200 AU, the gas surface density in the standard model is $\Sigma_g = 1.8 \
\rm{g \ cm}^{-2}$, which is optically thin in the vertical direction in
radio continuum region around 1 mm ($\tau_{\rm 1 mm} \sim 10^{-2}$).
Therefore, the density decline will be observed as a decrease in the
optical depth.

The disk is highly optically thick in the visible or near infrared light
($\tau_{\rm vis} \sim 200$ at 200 AU).
Thus, the density decrease is difficult to examine directly using
visible or near-infrared observations.
If the gap region is shadowed by the inner disk, or if the edge
of the outer disk is illuminated by the central star,
the variation in the surface irradiation will probably be observed.
There are a few Herbig Ae/Be stars that show gaps in the scattered light
images of their disks.
The disk around HD 163296 has a dark lane at 325 AU (Grady et al. 2000). 
HD 100546 also has a disk with a gap at 250 AU (Grady et al. 2001).
In order to make direct comparisons with these observations, we need
further modeling that takes account of the time evolution of the dust
density and that incorporates simulated scattered light images.

In this paper, we discussed gap formation through the surface stripping
caused by radiation pressure.
It is useful to discuss how to distinguish this mechanism from the other gap
formation processes such as the planetary perturbations (Lin \&
Papaloizou 1986; Takeuchi, Miyama \& Lin 1996).
Obviously, the surface stripping itself cannot generate non-axisymmetric
features, whereas an embedded planet can.
In gas-free disks, a planet forms characteristic disk structures that
arises from orbital resonances (Roques et al. 1994; Dermott et al. 1994;
Liou \& Zook 1999; Wyatt et al. 1999; Ozernoy et al. 2000; Wilner et
al. 2002; Quillen \& Thorndike 2002), and in gas disks the planet excites
spiral density waves (Miki 1982; Sekiya, Miyama, \& Hayashi 1987; Bryden
et al. 1999; Kley 1999; Lubow, Seibert, \& Artymowicz 1999; Miyoshi et
al. 1999; Tanigawa \& Watanabe 2002). 
The detection of such non-axisymmetric features would rule out surface
stripping as a gap formation mechanism.
The other notable feature of surface stripping is that it does not
affect the gas density structure.
The gas disks still have monotonic density structure even after the dust
disks have formed gaps.
Therefore, detection of gaps in gas disks also rules out surface
stripping.
In order to explain the cause of the gaps around HD 163296 and HD 100546,
it is important to determine with future observations whether their gas
disks have gaps or not.

\subsection{Summary}

Dust particles that are directly irradiated by the star move outward
under the combined action of gas drag and radiation pressure.
In this paper, the surface outflow of particles in optically thick disks
is studied.
Our results are as follows.

1. Particles in the surface layer, which are directly exposed to the
starlight, move outward.
In a disk resembling the minimum-mass-solar-nebula, the height of the
surface layer is about three scale heights of the gas disk 
and the mass is $10^{-6}$ times the total dust mass at 1 AU and
$10^{-4}$ times the total dust mass at 100 AU.
For this case, the mass flux of the surface outflow is much smaller than
the inflow mass flux inside the disk, and the surface outflow does not
contribute significantly to the dust disk evolution.

2. The surface outflow dominates the inflow in the following cases:
($a$) If the luminosity-mass ratio of the central star is 15 times
larger than the solar value, the velocity of the surface outflow
is large enough at the outer part of the disk ($r \ga 200$ AU).
($b$) If the mass of the gas disk is as small as $10^{-2}$ times the
minimum-mass disk ($\sim 100 M_{\earth}$ inside 100 AU), the mass ratio
of the surface layer to the whole dust disk is large enough at a
few tens AU.
In such cases, the surface outflow induces formation of gaps or rings.

3. The surface outflow is mainly composed of particles with size of
order $\sim 0.1 \ \micron$.
As the particles grow, the largest particles around the
midplane move inward more rapidly, while the outflow flux at the surface
remains at a near-constant value. This results in a diminishing
importance of the surface outflow for the overall disk evolution.

\acknowledgements

We would like to thank the referee for helpful comments.
We also thank Greg Laughlin for his careful reading the manuscript and
correcting grammar.
This work was supported in part by an NSF grant AST 99 87417 and in part
by a special NASA astrophysical theory program that supports a joint
Center for Star Formation Studies at UC Berkeley, NASA-Ames Research
Center, and UC Santa Cruz.
This work is also supported by NASA NAG5-10612 through its Origin
program and by JPL 1228184 through its SIM program.



\clearpage

\begin{figure}
\epsscale{1.0}
\plotone{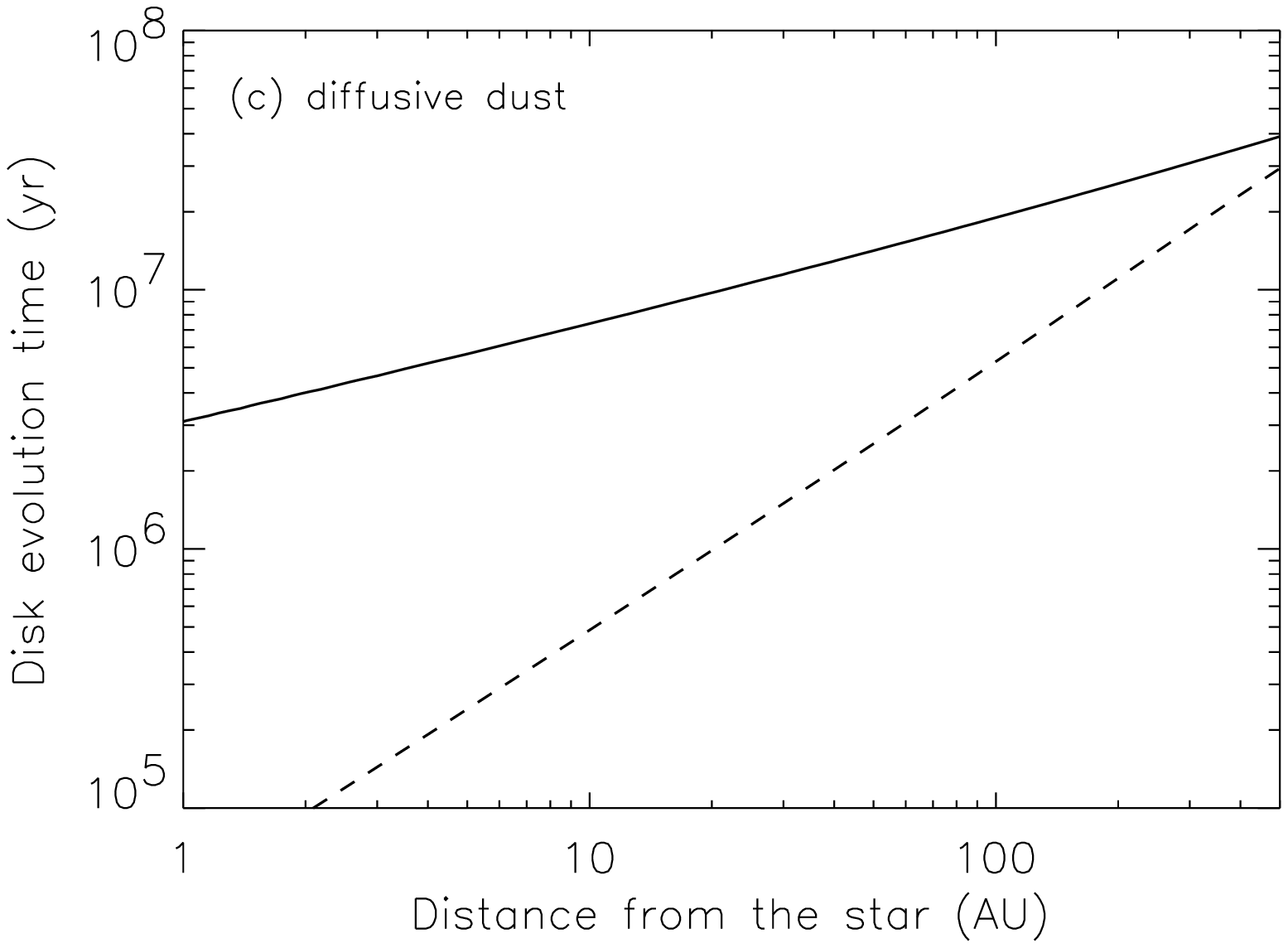}
Fig. 6 --- continued
\end{figure}

\begin{figure}
\epsscale{1.0}
\plotone{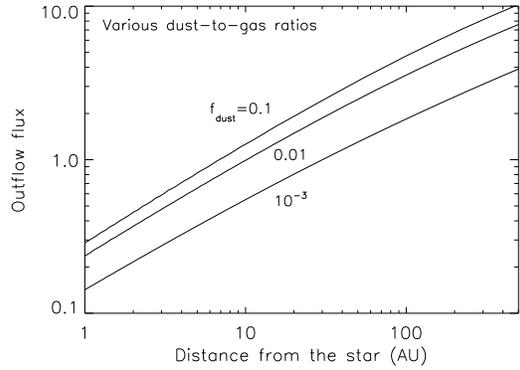}
\caption{
Surface dust flux for various dust-to-gas ratios, $f_{\rm dust} = 0.1$,
0.01, and $10^{-3}$ from the upper line. The flux is normalized by the
value of the standard model ($f_{\rm dust} = 0.01$) at 10 AU.
\label{fig:flux}
}
\end{figure}

\end{document}